\documentclass{article}
\usepackage{spconf,amsmath,graphicx,hyperref,comment}

\usepackage{adjustbox}
\usepackage{caption}
\usepackage{multirow}
\usepackage{booktabs}
\usepackage{array}
\usepackage{hhline}

\title{DistilMOS: Layer-wise Self-distillation for \\ Self-Supervised Learning Model-based MOS Prediction}
\name{
Jianing Yang, Wataru Nakata, Yuki Saito, and Hiroshi Saruwatari}
\address{The University of Tokyo, Japan.}
\begin{document}
\ninept

\maketitle

\begin{abstract}
With the advancement of self-supervised learning (SSL), fine-tuning pretrained SSL models for mean opinion score (MOS) prediction has achieved state-of-the-art performance. However, during fine-tuning, these SSL-based MOS prediction models often suffer from catastrophic forgetting of the pretrained knowledge and tend to overfit the training set, resulting in poor generalization performance.
In this study, we propose DistilMOS, a novel method that learns to predict not only MOS but also token IDs obtained by clustering the hidden representations of each layer in the pretrained SSL model.
These layer-wise token targets serve as self-distillation signals that enables the MOS prediction model to extract rich internal knowledge from SSL models, enhancing both prediction accuracy and generalization capability.
Experimental evaluations demonstrate that our method significantly outperforms standard SSL-based MOS prediction models on both in-domain and out-of-domain evaluations, verifying the effectiveness and practicality of the proposed method\footnote{Code: \url{https://github.com/BaleYang/DistilMOS}}.
\end{abstract}


%

\vspace{-8pt}
\section{Introduction}
\label{sec:intro}
\vspace{-5pt}

\begin{figure*}[t]
    \centering
    \includegraphics[width=0.89\textwidth]{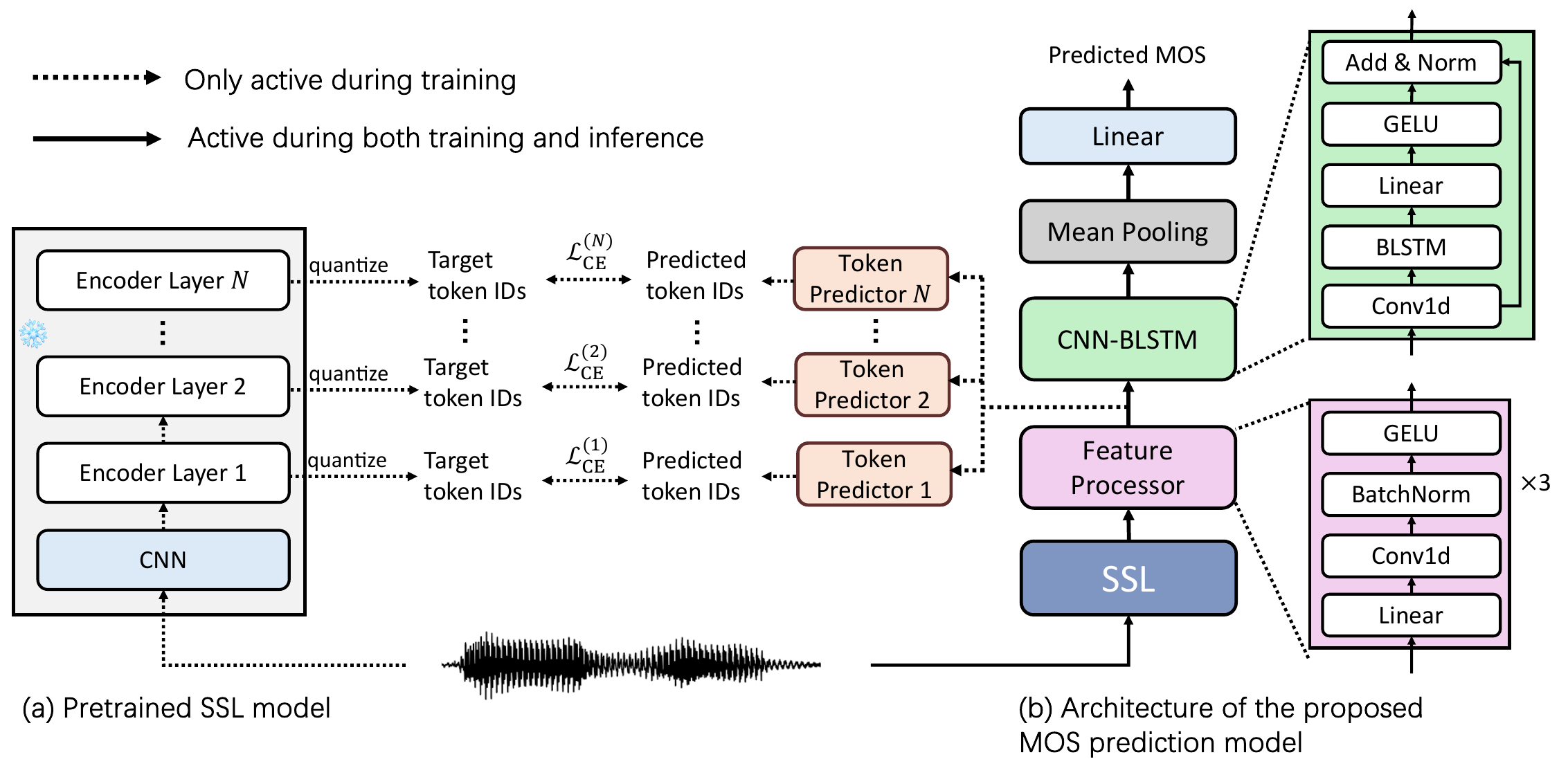}
    \vspace{-3pt}
    
    \caption{\normalfont Overview of the proposed method. (a) Pretrained SSL model: frame-level representations from each encoder layer are quantized via $k$-means clustering to generate discrete token IDs. (b) Proposed MOS prediction model architecture: during training, the model is jointly optimized with two objectives, MOS regression and token prediction for self-distillation, where token predictors learn to reconstruct the quantized token IDs from each SSL layer.}
    \label{fig:architecture}
\end{figure*}

Subjective speech quality assessment is a crucial means of evaluating the performance of text-to-speech (TTS) and voice conversion (VC) systems. Among such speech quality assessment, the mean opinion score (MOS) test evaluates speech naturalness based on the average of multiple human listeners' ratings. However, manual MOS testing is time-consuming and costly, motivating the development of automatic MOS prediction models based on deep neural networks (DNNs). For instance, MOSNet~\cite{mosnet} adopts a CNN-BLSTM architecture to predict MOS from input speech at the utterance level and achieved high correlation with human ratings on the Voice Conversion Challenge 2018 dataset~\cite{lorenzo2018voice}. Subsequent works considered listener bias during MOS test. For example, MBNet~\cite{mbnet} and LDNet~\cite{huang2022ldnet} take individual listener ratings as additional inputs to model inter-listener variability, thereby improving prediction accuracy.
Recently, with the emergence of large-scale self-supervised learning (SSL) models pretrained on speech~\cite{baevski2020wav2vec,hsu2021hubert,chen2022wavlm}, leveraging feature representations from hidden layers of these models for MOS prediction has become a mainstream method~\cite{SSLMOS}. SSL-based models have demonstrated strong performance on public datasets, and in the VoiceMOS Challenge 2022~\cite{huang2022voicemos}, most top-performing systems were built upon or extended from the SSL-MOS baseline~\cite{saeki2022utmos, tseng2022ddos, tian2022transfer, yang2022fusion}.

Despite achieving impressive accuracy, existing SSL-driven MOS prediction models still face several limitations. For instance, fine-tuning SSL models solely with MOS regression may lead to catastrophic forgetting~\cite{chen2022continual} of previously learned knowledge, thus impairing the model’s generalization capability~\cite{zhang2023you}. It has been reported that SSL-based MOS prediction models suffer a significant performance drop in zero-shot evaluations across different datasets~\cite{SSLMOS}. Many studies have shown that different layers of SSL models capture distinct levels of speech information. That is, lower layers tend to encode acoustic features such as speaker and prosodic cues, while higher layers capture more abstract information like lexical and semantic content~\cite{chen2022wavlm, pasad2021layer, de2024layer, choi2024self}. These hierarchical features are all critical in human perception of speech quality~\cite{cooper2024review, vioni2023investigating}. Hence, an urgent problem is how to effectively utilize the multi-level representations of SSL models to improve model robustness and generalization across domains in MOS prediction.

To address this issue, we propose a novel MOS prediction method based on layer-wise self-distillation. Our core idea is to incorporate a self-distillation mechanism, where the MOS prediction model learns not only to predict MOS but also to reconstruct the latent feature patterns from the internal layers of the pretrained SSL model. Concretely, we apply $k$-means clustering~\cite{hartigan1979algorithm} to the frame-level outputs of each SSL layer to generate discrete token ID sequences. During model training, we introduce token predictors, which are supervised to classify the corresponding token sequences from each layer. This multi-task training enables the MOS prediction model to focus on both the acoustic and semantic information encoded at different levels of the SSL model without requiring additional data.
Experiments on the BVCC dataset~\cite{SSLMOS} demonstrate that our method achieves better performance compared to baseline. Furthermore, it also performs well in zero-shot prediction on the SOMOS dataset~\cite{maniati2022somos}, confirming the effectiveness and robustness of our method for MOS prediction tasks.

\vspace{-5pt}
\section{DistilMOS}
\vspace{-5pt}
\subsection{Overall Architecture}
\vspace{-3pt}

The overall architecture of the proposed MOS prediction model are illustrated in Fig.~\ref{fig:architecture}(b). As in prior work, we first extract frame-level representations from all Transformer blocks of an SSL model initialized with pretrained weights and fine-tuned during training, and then aggregate them by a learnable layer-wise weighted sum~\cite{baba2024t05}. These features are then passed through a \textit{Feature Processor}, which consists of three stacked blocks, each consisting of a linear layer, a 1D convolution (Conv1D), batch normalization, and a GELU activation~\cite{hendrycks2016gaussian}. This module is designed to capture local contextual information along the sequence. Because the SSL representations are high-dimensional, we insert a linear projector between the SSL encoder and the Feature Processor to reduce the dimensionality to 256.

The output of the Feature Processor is fed into a \textit{CNN-BLSTM} module. The CNN-BLSTM structure begins with a Conv1D layer, followed by a BLSTM~\cite{lstm} to model temporal dependencies. The forward and backward sequence embeddings are concatenated and projected back to the original dimension via a linear layer, then activated by GELU. Finally, a residual connection with the initial Conv1D output is applied, followed by a layer normalization~\cite{ba2016layer_corr} operation, yielding the final output of the CNN-BLSTM.

To predict the utterance-level MOS, we apply mean pooling to the sequential output from the CNN-BLSTM, followed by a linear projection. The MOS prediction is optimized with a mean squared error (MSE) loss between the ground-truth MOS $s$ and predicted MOS $\hat{s}$:
\begin{equation}
\mathcal{L}_{\text{MOS}} = (\hat{s} - s)^2.
\end{equation}

\vspace{-8pt}
\subsection{Self-Distillation from Discretized SSL Features}
\vspace{-3pt}

To incorporate richer internal knowledge from the pretrained SSL model, we adopt a token prediction strategy. As illustrated in Fig.~\ref{fig:architecture}(a), for each layer of the pretrained SSL model, we apply $k$-means clustering to the frame-level representations to generate discrete token sequences for each utterance. These cluster assignments serve as pseudo labels representing difference speech features of each layer.

During training, the output from the Feature Processor is fed into $N$ parallel \textit{Token Predictors} (one for each SSL layer), where each predictor consists of a 3-layer multi-layer perceptron (MLP) with GELU activations. These modules are trained to classify the token IDs obtained from the corresponding SSL layer.

The token classification for the $n$-th layer is optimized to minimize a cross-entropy loss:
\begin{equation}
\mathcal{L}_{\text{CE}}^{(n)} = -\frac{1}{L} \sum_{i=1}^{L} \log p_i^{(n)},
\end{equation}
where $L$ denotes the sequence length and $p_i^{(n)}$ is the predicted probability of the correct token at position $i$ for layer $n$.

The motivation behind this self-distillation strategy is that human MOS perception depends on various levels of speech features, such as prosody, speaking rate, and intelligibility~\cite{cooper2024review}. The token prediction task implicitly encourages the MOS prediction model to learn such acoustic and semantic features distributed across different SSL layers~\cite{pasad2021layer}. When these representations are fused into the MOS prediction head, they provide a stronger basis for subjective quality estimation.

Moreover, since the token labels are derived from the pretrained SSL model, no additional human-annotated data is required. This makes our method efficient and scalable. The token prediction branch can be discarded during inference, incurring no extra computational cost. This design also mitigates the risk of in-domain overfitting caused by direct MOS-only optimization and improves generalization to unseen domains.


Unlike conventional distillation methods such as DistilHuBERT~\cite{chang2022distilhubert}, which aim to predict the exact embeddings of each SSL layer, our method instead focuses on learning essential representations for MOS prediction while discarding irrelevant details. This design prioritizes task-relevant abstraction rather than full feature preservation.

\vspace{-5pt}
\subsection{Training Objective}
\vspace{-3pt}

The final training objective combines MOS regression and token classification across $N$ SSL layers:
\begin{equation}
\mathcal{L} = \mathcal{L}_{\text{MOS}} + \alpha \cdot \frac{1}{N} \sum_{n=1}^{N} \mathcal{L}_{\text{CE}}^{(n)},
\end{equation}
where $\alpha$ is a weighting coefficient that balances the contribution of the token-level self-distillation loss.

\vspace{-8pt}
\section{Experimental evaluation}
\label{sec:experiment}
\vspace{-5pt}

\begin{table*}[t]
\caption{Experimental results on BVCC (in-domain) and SOMOS (zero-shot) datasets with different SSL backbones. \textbf{Bold} value indicates the best result for each SSL backbone.}

\resizebox{\textwidth}{!}{
\centering
\renewcommand{\arraystretch}{1.05}
\begin{tabular}{|c|l|cccc|cccc|cccc|}
\hhline{|-| - |----|----|----|}
\multirow{3}{*}{\centering\textbf{SSL backbone}} &
\multirow{3}{*}{\centering\textbf{\;\;\;\;\;\;\;\;\;\;\;Model}} &
\multicolumn{8}{c|}{\textbf{BVCC}} &
\multicolumn{4}{c|}{\textbf{SOMOS}} \\
\hhline{|~|~|----|----|----|}
 &  &
\multicolumn{4}{c|}{\textbf{Utterance-level}} &
\multicolumn{4}{c|}{\textbf{System-level}} &
\multicolumn{4}{c|}{\textbf{Utterance-level}} \\
\hhline{|~|~|----|----|----|}
 & & LCC & SRCC & KTAU & MSE & LCC & SRCC & KTAU & MSE & LCC & SRCC & KTAU & MSE \\
\hhline{|=|=|====|====|====|}
\multirow{4}{*}{Wav2Vec2 Base}
  & SSL-MOS               & 0.869 & 0.867 & 0.687 & 0.472 & 0.926 & 0.923 & 0.769 & 0.301 & 0.320 & 0.310 & 0.210 & \textbf{0.654} \\
  & \textbf{DistilMOS (ours)} & \textbf{0.884} & \textbf{0.883} & \textbf{0.711} & \textbf{0.186} & \textbf{0.934} & \textbf{0.934} & \textbf{0.782} & 0.098 & \textbf{0.344} & \textbf{0.318} & \textbf{0.217} & 0.960 \\
  & w/o token prediction  & 0.872 & 0.871 & 0.695 & 0.231 & 0.928 & 0.927 & 0.774 & 0.143 & 0.316 & 0.305 & 0.207 & 1.403 \\
  & MSE distillation & \textbf{0.884} & 0.881 & 0.709 & 0.187 & 0.932 & 0.930 & 0.777 & \textbf{0.089} & 0.277 & 0.255 & 0.173 & 0.940 \\
\hhline{|=|=|====|====|====|}
\multirow{4}{*}{WavLM Base}
  & SSL-MOS               & 0.869 & 0.870 & 0.694 & 0.354 & 0.917 & 0.916 & 0.763 & 0.210 & 0.353 & 0.336 & 0.230 & 0.973 \\
  & \textbf{DistilMOS (ours)} & \textbf{0.886} & \textbf{0.884} & \textbf{0.712} & \textbf{0.190} & \textbf{0.931} & \textbf{0.928} & \textbf{0.778} & \textbf{0.087} & \textbf{0.396} & \textbf{0.375} & \textbf{0.257} & 1.051 \\
  & w/o token prediction  & 0.881 & 0.882 & 0.711 & \textbf{0.190} & 0.925 & 0.922 & 0.768 & 0.094 & 0.312 & 0.297 & 0.203 & \textbf{0.796} \\
  & MSE distillation & 0.874	& 0.871	& 0.695	& 0.214	& 0.923	& 0.919	& 0.763	& 0.098	& 0.383	& 0.362	& 0.248 & 0.935\\

\hhline{|=|=|====|====|====|}
\multirow{4}{*}{WavLM Base+}
  & SSL-MOS               & 0.862 & 0.870 & 0.691 & 0.281 & 0.917 & 0.927 & 0.779 & 0.142 & 0.315 & 0.287 & 0.196 & \textbf{0.317} \\
  & \textbf{DistilMOS (ours)} & \textbf{0.885} & \textbf{0.885} & \textbf{0.713} & \textbf{0.191} & \textbf{0.936} & \textbf{0.932} & \textbf{0.782} & \textbf{0.083} & \textbf{0.396} & \textbf{0.370} & \textbf{0.253} & 0.998 \\
  & w/o token prediction  & 0.879 & 0.878 & 0.705 & 0.193 & 0.931 & 0.927 & 0.775 & 0.090 & 0.320 & 0.297 & 0.202 & 0.748 \\
  & MSE distillation & 0.882	& 0.881	& 0.709	& 0.193	& 0.927	& 0.922	& 0.773	& 0.091	& 0.277	& 0.360	& 0.225	& 0.724 \\

\hhline{|=|=|====|====|====|}
\multirow{4}{*}{WavLM Large}
  & SSL-MOS               & 0.884 & 0.881 & 0.708 & 0.195 & 0.934 & 0.930 & 0.784 & 0.104 & 0.197 & 0.188 & 0.127 & 1.829 \\
  & \textbf{DistilMOS (ours)} & \textbf{0.891} & \textbf{0.889} & \textbf{0.720} & 0.188 & \textbf{0.939} & \textbf{0.936} & \textbf{0.791} & 0.105 & \textbf{0.404} & \textbf{0.387} & \textbf{0.264} & 0.965 \\
  & w/o token prediction  & 0.888 & 0.887 & 0.718 & \textbf{0.180} & 0.934 & 0.932 & 0.782 & \textbf{0.088} & 0.393 & 0.375 & 0.256 & \textbf{0.823} \\
  & MSE distillation & 0.887	& 0.886	& 0.715	& 0.184	& 0.933	& 0.930	& 0.779	& 0.097	& 0.387	& 0.370	& 0.252	& 1.200 \\

\hhline{|-| - |----|----|----|}
\end{tabular}
}
\label{table:result}
\end{table*}

\subsection{Datasets}
\vspace{-3pt}

Our experiments were conducted on the BVCC dataset~\cite{SSLMOS}, the official corpus for the main track of the VoiceMOS Challenge 2022~\cite{huang2022voicemos}. BVCC consists of 7,106 English synthetic speech utterances generated from 187 different systems across multiple Blizzard Challenges~\cite{blizzard2009} and Voice Conversion Challenges~\cite{lorenzo2018voice}. Each utterance was rated by 8 listeners for naturalness on a 1–5 scale. We followed the official split, using 4,974 utterances for training, 1,066 for validation, and 1,066 for testing.

To further examine the robustness of our MOS prediction model, we adopted a zero-shot evaluation setting on the SOMOS dataset (without fine-tuning)~\cite{maniati2022somos}. SOMOS is a large-scale open-source corpus for subjective naturalness assessment of neural TTS systems. It contains 20,000 synthetic utterances generated from 200 acoustic models trained on the LJSpeech dataset~\cite{ljspeech17} (mainly Tacotron-based~\cite{wang2017tacotron}), along with 100 natural utterances. Each utterance was rated by at least 17 different listeners on a 1–5 scale.

\begin{figure*}[t]
    \centering
    \includegraphics[width=0.9\textwidth]{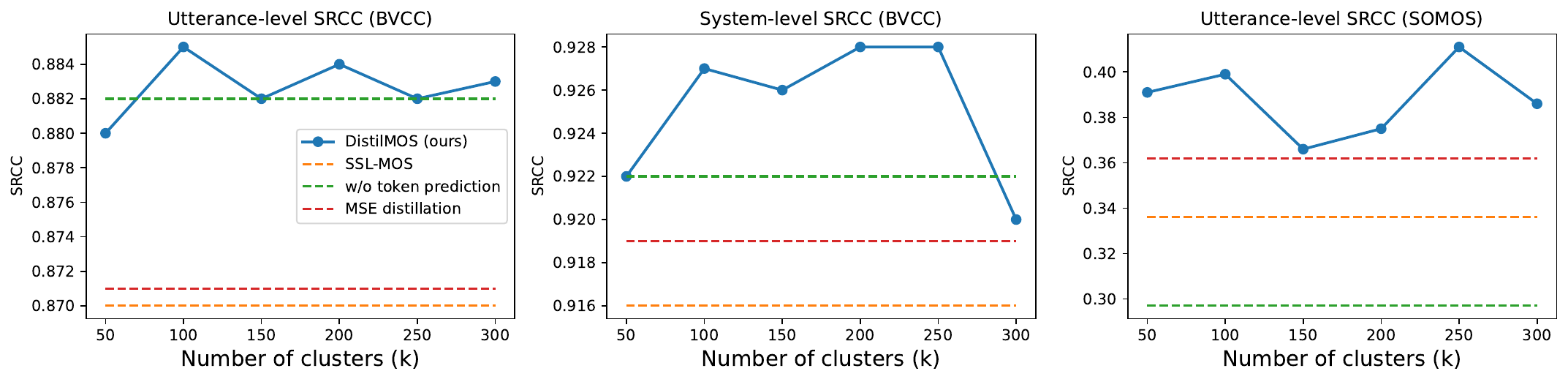}
    \caption{\normalfont Effect of the number of clusters $k$ on SRCC performance. Left: utterance-level SRCC on BVCC (in-domain) dataset. Middle: system-level SRCC on BVCC dataset. Right: utterance-level SRCC on SOMOS (zero-shot) dataset.}
    \label{fig:cluster}
\end{figure*}

\vspace{-5pt}
\subsection{Evaluation Metrics}
\vspace{-3pt}
For in-domain experiments on the BVCC dataset, we report results at both the utterance level and the system level using four metrics: MSE, linear correlation coefficient (LCC), Spearman rank correlation coefficient (SRCC), and Kendall’s $\tau$ (KTAU). Higher correlation coefficients indicate better alignment between predicted and human ratings, while lower MSE indicates smaller prediction errors.
For out-of-domain zero-shot evaluation on SOMOS, we report only utterance-level metrics, which impose stricter requirements on generalization to unseen utterances and ranking consistency.

\vspace{-5pt}
\subsection{Implementation Details}
\vspace{-3pt}
To verify the general applicability of our method, we conducted experiments using four different SSL models: Wav2Vec2 Base~\cite{baevski2020wav2vec}, WavLM Base, WavLM Base+, and WavLM Large~\cite{chen2022wavlm}. Wav2Vec2 Base and WavLM Base are of similar size and trained on the same datasets, but differ in pretraining objectives: Wav2Vec2 uses contrastive learning, whereas WavLM Base follows HuBERT-style masked token prediction~\cite{hsu2021hubert}.
WavLM Base and WavLM Base+ share the same architecture, but WavLM Base+ was pretrained on a larger dataset, thus possessing richer knowledge. WavLM Base+ and WavLM Large were trained on the same dataset, but differ in model size.

All linear, Conv1D, and BLSTM modules within the Feature Processor and CNN-BLSTM were set to 256 dimensions.
For clustering, we applied $k$-means clustering ($k = 200$) to each SSL layer’s outputs on the BVCC training set to obtain token IDs for each utterance. To accelerate centroid estimation, we used MiniBatchKMeans~\cite{sculley2010web} with the online partial fit routine, updating clusters with batch size 64. We empirically set the token distillation weight to $\alpha = 0.1$. Training was performed for 10,000 steps with a batch size of 32 on a single NVIDIA GH200 (120\,GB) GPU. Checkpoints were saved every 1,000 steps, and the model with the highest utterance-level SRCC on the validation set was selected for evaluation on the test set.

We used the AdamW optimizer~\cite{loshchilov2017decoupled} with a OneCycleLR scheduler~\cite{smith2019super}. Hyperparameters were set as: learning rate = 1e-4, betas = (0.9, 0.98), weight decay = 1e-4, and gradient clipping threshold = 10.0.  

\vspace{-5pt}
\subsection{Baseline and Ablation Studies}
\vspace{-3pt}
Although our proposed method can be applied to any MOS prediction models that use pretrained SSL models for feature extraction, we adopted SSL-MOS~\cite{SSLMOS} as the baseline for simplicity. To eliminate potential bias introduced by differences in SSL backbones, we fine-tuned each of the four SSL models described in Section 3.3 and compared their performance against our proposed method initialized with the same backbones.

To further validate the effectiveness of our method, we conducted ablation studies. Specifically, for each configuration, we trained a variant without token prediction, i.e., using only the proposed architecture optimized with the MSE loss (Eq.~(1)). In addition, to demonstrate the effectiveness of predicting discrete tokens, we performed another variant where, instead of token prediction, the model was equipped with $N$ prediction heads (one per SSL layer) to predict the original SSL embeddings directly, and optimized with an MSE loss.

We further examine the sensitivity of our method to the number of clusters used for SSL layer discretization. Using WavLM Base as the backbone, we vary $k \in \{50, 100, 150, 200, 250, 300\}$ and report the corresponding SRCC on both BVCC (in-domain) and SOMOS (zero-shot) dataset.

\vspace{-5pt}
\subsection{Results}
\vspace{-3pt}

The experimental results in Table~\ref{table:result} show that our method surpasses SSL-MOS across all backbones and metrics on BVCC, with the large gains on system-level and utterance-level SRCC. On SOMOS (zero-shot), it improves all correlation metrics, indicating better in-domain performance and stronger robustness to unseen conditions. The relatively high MSE on SOMOS is likely caused by cross-dataset score calibration differences, under which correlation-based metrics better reflect perceptual consistency than absolute error.

The results of the ablation studies further validate the contribution of token prediction. Removing token prediction and training the model solely with the MOS regression loss (w/o token prediction) led to noticeable performance degradation.
Moreover, replacing token prediction with prediction heads that regress the SSL embeddings directly (MSE distillation) also resulted in performance declines, with all correlation-based metrics falling significantly behind the token prediction method. We speculate that this is because MSE-based distillation makes the MOS prediction model overfit the pretrained SSL features, leading to poor performance.

The results of varying the number of clusters are shown in Fig.~\ref{fig:cluster}. We observe that the cluster size does not significantly affect performance, as consistently strong results are maintained across all values of $k$. Notably, in the zero-shot SOMOS evaluation, all cluster configurations surpass both the baseline and ablation variants, further demonstrating the robustness and generalization capability of our proposed method.

\subsection{Analysis of Learned Representations}
To examine whether our token-prediction self-distillation method enables the MOS prediction model to capture rich speech features from the pretrained SSL backbone, we conducted canonical correlation analysis (CCA)~\cite{hotelling1992relations}. Specifically, for both MOS prediction model of the proposed method and the ablation baselines, we computed the CCA between the average-pooled outputs of the Feature Processor and the average-pooled representations from each Transformer layer of the pretrained SSL model.
We additionally performed the same analysis on a randomly initialized MOS prediction model
 of the proposed model to assess whether architectural biases affect the results. For the SSL-MOS baseline, we instead computed the CCA between the average-pooled output of its final Transformer layer with the average-pooled representations from each Transformer layer of the pretrained SSL model. All experiments were conducted using WavLM Base as the backbone. We computed the CCA following~\cite{pasad2023comparative}\footnote{\url{https://github.com/ankitapasad/layerwise-analysis/}} on BVCC test set.

\begin{figure}[t]
    \centering
    \includegraphics[width=1\linewidth]{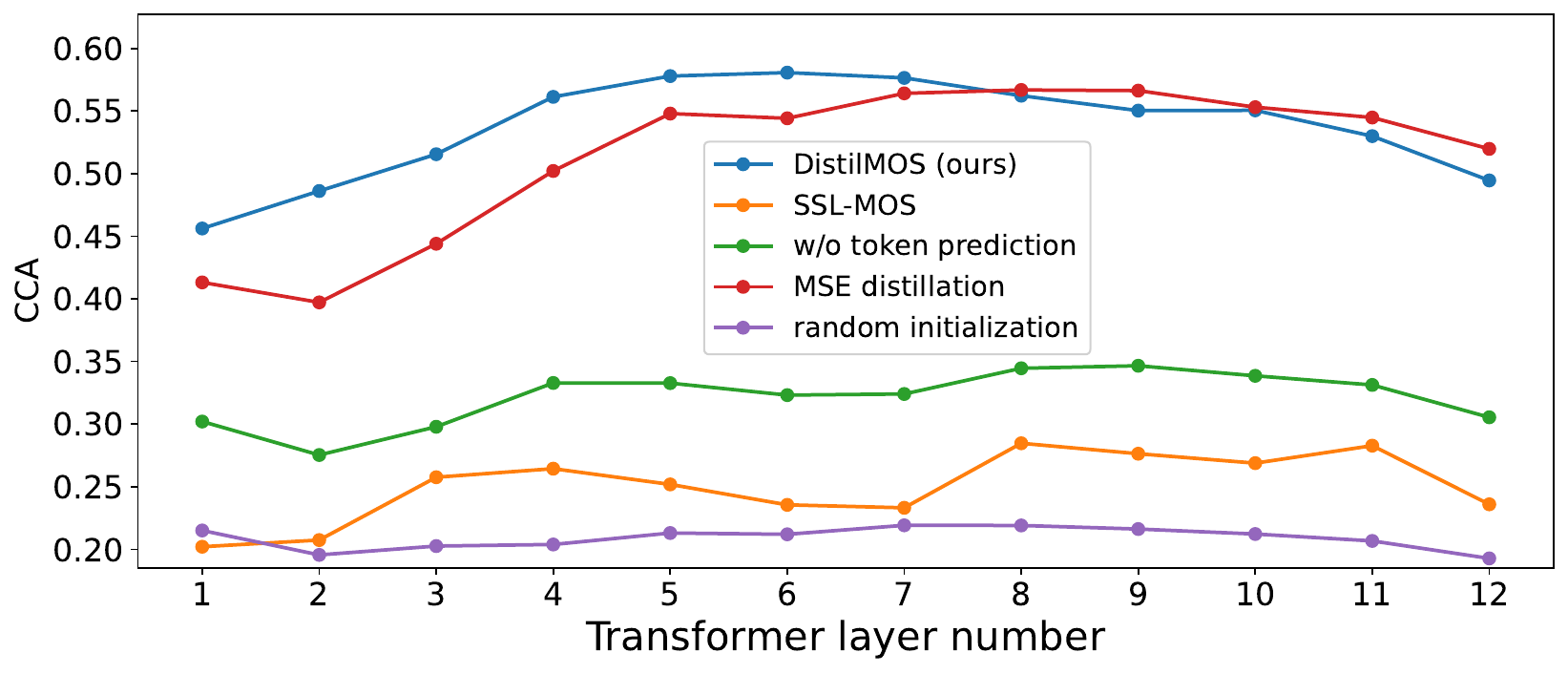}
    \caption{\normalfont CCA curves between MOS prediction model representations and pretrained SSL layer outputs.}
    \vspace{-5pt}
    \label{fig:layer_analysis}

\end{figure}

As shown in Fig.~\ref{fig:layer_analysis}, both self-distillation method (DistilMOS and MSE distillation) maintain consistently high CCA values across layers, indicating that self-distillation successfully encourages the MOS prediction model to learn rich speech features present from the pretrained SSL model. In contrast, the MOS-only fine-tuned model (SSL-MOS and w/o token prediction) exhibits correlations close to random initialization, suggesting severe catastrophic forgetting. This finding provides further evidence that catastrophic forgetting undermines generalization in zero-shot scenarios.

\vspace{-8pt}
\section{Conclusion}
\vspace{-5pt}
In this paper, we proposed a novel method that enables the model to predict token IDs derived from clustering its own intermediate representations, thereby extracting and leveraging its internal knowledge. This self-distillation strategy effectively enhances both performance and generalization capability. Our method achieves superior results compared to baselines on both BVCC (in-domain) and SOMOS (zero-shot) datasets, and the ablation studies further validate its feasibility and robustness.

Since BVCC is a relatively small dataset, the clustering results may exhibit mismatches when applied to other datasets. As part of future work, we plan to extend this method to larger or combined datasets to further refine the method and develop a MOS prediction model with stronger generalization ability.

{\bf Acknowledgements:}
The work was supported by JST Moonshot Grant Number JPMJMS2011 and based on results obtained from a project outsourced by the New Energy and Industrial Technology Development Organization (NEDO).

\bibliographystyle{IEEEbib}

\footnotesize
\bibliography{refs}

@inproceedings{SSLMOS,
  title={{Generalization ability of MOS prediction networks}},
  author={Cooper, Erica and Huang, Wen-Chin and Toda, Tomoki and Yamagishi, Junichi},
  booktitle={Proc. ICASSP},
  pages={8442--8446},
  year={2022},
}

@inproceedings{mosnet,
  author    = {Chen-Chou Lo and Szu-Wei Fu and Wen-Chin Huang and Xin Wang and Junichi Yamagishi and Yu Tsao and Hsin-Min Wang},
  title     = {{MOSNet: Deep Learning based Objective Assessment for Voice Conversion}},
  booktitle = {Proc. Interspeech},
  year      = {2019},
  pages     = {1541--1545},
  doi       = {10.21437/Interspeech.2019-2003},
  url       = {https://www.isca-speech.org/archive/interspeech_2019/lo19_interspeech.html}
}

@inproceedings{lorenzo2018voice,
  title={{The Voice Conversion Challenge 2018: Promoting development of parallel and nonparallel methods}},
  author={Lorenzo-Trueba, Jaime and Yamagishi, Junichi and Toda, Tomoki and Saito, Daisuke and Villavicencio, Fernando and Kinnunen, Tomi and Ling, Zhenhua},
  booktitle={Proc. Speaker Odyssey},
  pages={195--202},
  year={2018},
}

@inproceedings{mbnet,
  title={{MBNet: MOS prediction for synthesized speech with mean-bias network}},
  author={Leng, Yichong and Tan, Xu and Zhao, Sheng and Soong, Frank and Li, Xiang-Yang and Qin, Tao},
  booktitle={Proc. ICASSP},
  pages={391--395},
  year={2021},
}

@inproceedings{huang2022ldnet,
  title={{LDNet: Unified listener dependent modeling in MOS prediction for synthetic speech}},
  author={Huang, Wen-Chin and Cooper, Erica and Yamagishi, Junichi and Toda, Tomoki},
  booktitle={Proc. ICASSP},
  pages={896--900},
  year={2022},
}

@inproceedings{baevski2020wav2vec,
  title={{wav2vec 2.0: A framework for self-supervised learning of speech representations}},
  author={Baevski, Alexei and Zhou, Yuhao and Mohamed, Abdelrahman and Auli, Michael},
  booktitle={Proc. NeurIPS},
  pages={12449--12460},
  year={2020}
}

@article{hsu2021hubert,
  title={{HuBERT: Self-supervised speech representation learning by masked prediction of hidden units}},
  author={Hsu, Wei-Ning and Bolte, Benjamin and Tsai, Yao-Hung Hubert and Lakhotia, Kushal and Salakhutdinov, Ruslan and Mohamed, Abdelrahman},
  journal={IEEE/ACM transactions on audio, speech, and language processing},
  volume={29},
  pages={3451--3460},
  year={2021},
}

@article{chen2022wavlm,
  title={{WavLM: Large-scale self-supervised pre-training for full stack speech processing}},
  author={Chen, Sanyuan and Wang, Chengyi and Chen, Zhengyang and Wu, Yu and Liu, Shujie and Chen, Zhuo and Li, Jinyu and Kanda, Naoyuki and Yoshioka, Takuya and Xiao, Xiong and others},
  journal={IEEE Journal of Selected Topics in Signal Processing},
  volume={16},
  number={6},
  pages={1505--1518},
  year={2022},
}

@inproceedings{tseng2022ddos,
  title={{DDOS: A MOS prediction framework utilizing domain adaptive pre-training and distribution of opinion scores}},
  author={Tseng, Wei-Cheng and Kao, Wei-Tsung and Lee, Hung-yi},
  booktitle={Proc. Interspeech},
  pages={4541--4545},
  year={2022}
}

@inproceedings{saeki2022utmos,
  author    = {Takaaki Saeki and Detai Xin and Wataru Nakata and Tomoki Koriyama and Shinnosuke Takamichi and Hiroshi Saruwatari},
  title     = {{UTMOS: UTokyo-SaruLab System for VoiceMOS Challenge 2022}},
  booktitle = {Proc. Interspeech},
  year      = {2022},
  pages     = {4521--4525},
}

@inproceedings{huang2022voicemos,
  title={{The VoiceMOS Challenge 2022}},
  author={Huang, Wen-Chin and Cooper, Erica and Tsao, Yu and Wang, Hsin-Min and Toda, Tomoki and Yamagishi, Junichi},
  booktitle={Proc. Interspeech},
  pages={4536--4540},
  year={2022}
}

@inproceedings{yang2022fusion,
  author    = {Zhengdong Yang and Wangjin Zhou and Chenhui Chu and Sheng Li and Raj Dabre and Raphael Rubino and Yi Zhao},
  title     = {{Fusion of self-supervised learned models for MOS prediction}},
  booktitle = {Proc. Interspeech},
  pages     = {5443--5447},
  year      = {2022},
}

@inproceedings{tian2022transfer,
  title={{A transfer and multi-task learning based approach for MOS prediction}},
  author={Tian, Xiaohai and Fu, Kaiqi and Gao, Shaojun and Gu, Yiwei and Wang, Kai and Li, Wei and Ma, Zejun},
  booktitle={Proc. Interspeech},
  pages={5438--5442},
  year={2022}
}

@inproceedings{pasad2021layer,
  title={{Layer-wise analysis of a self-supervised speech representation model}},
  author={Pasad, Ankita and Chou, Ju-Chieh and Livescu, Karen},
  booktitle={Proc. ASRU},
  pages={914--921},
  year={2021},
}

@inproceedings{de2024layer,
  title={{A layer-wise analysis of Mandarin and English suprasegmentals in SSL speech models}},
  author={de la Fuente, Anton and Jurafsky, Dan},
  booktitle={Proc. Interspeech},
  pages={1290--1294},
  year={2024}
}

@inproceedings{choi2024self,
  title={{Self-Supervised Speech Representations are More Phonetic than Semantic}},
  author={Choi, Kwanghee and Pasad, Ankita and Nakamura, Tomohiko and Fukayama, Satoru and Livescu, Karen and Watanabe, Shinji},
  booktitle={Proc. Interspeech},
  pages={4578--4582},
  year={2024}
}

@incollection{chen2022continual,
  title={{Continual learning and catastrophic forgetting}},
  author={Chen, Zhiyuan and Liu, Bing},
  booktitle={Lifelong Machine Learning},
  pages={55--75},
  year={2022},
  publisher={Springer}
}

@inproceedings{zhang2023you,
  title={{Do you remember? Overcoming catastrophic forgetting for fake audio detection}},
  author={Zhang, Xiaohui and Yi, Jiangyan and Tao, Jianhua and Wang, Chenglong and Zhang, Chu Yuan},
  booktitle={Proc. ICML},
  pages={41819--41831},
  year={2023},
}

@inproceedings{maniati2022somos,
  title={{SOMOS: The Samsung Open MOS Dataset for the Evaluation of Neural Text-to-Speech Synthesis}},
  author={Maniati, Georgia and Vioni, Alexandra and Ellinas, Nikolaos and Nikitaras, Karolos and Klapsas, Konstantinos and Sung, June Sig and Jho, Gunu and Chalamandaris, Aimilios and Tsiakoulis, Pirros},
  booktitle={Proc. Interspeech},
  pages={2388--2392},
  year={2022}
}

@article{hartigan1979algorithm,
  title={{Algorithm AS 136: A k-means clustering algorithm}},
  author={Hartigan, John A and Wong, Manchek A},
  journal={Journal of the Royal Statistical Society. Series C (Applied Statistics)},
  volume={28},
  number={1},
  pages={100--108},
  year={1979},
  publisher={JSTOR}
}

@article{hendrycks2016gaussian,
  title={{Gaussian error linear units (GELUs)}},
  author={Hendrycks, Dan and Gimpel, Kevin},
  journal={arXiv preprint arXiv:1606.08415},
  year={2016}
}

@article{lstm,
  title={{Long short-term memory}},
  author={Hochreiter, Sepp and Schmidhuber, J{\"u}rgen},
  journal={Neural computation},
  volume={9},
  number={8},
  pages={1735--1780},
  year={1997},
  publisher={MIT press}
}

@article{cooper2024review,
  title={{A review on subjective and objective evaluation of synthetic speech}},
  author={Cooper, Erica and Huang, Wen-Chin and Tsao, Yu and Wang, Hsin-Min and Toda, Tomoki and Yamagishi, Junichi},
  journal={Acoustical Science and Technology},
  volume={45},
  number={4},
  pages={161--183},
  year={2024},
  publisher={ACOUSTICAL SOCIETY OF JAPAN}
}

@inproceedings{chang2022distilhubert,
  title={{DistilHuBERT: Speech representation learning by layer-wise distillation of hidden-unit BERT}},
  author={Chang, Heng-Jui and Yang, Shu-wen and Lee, Hung-yi},
  booktitle={Proc. ICASSP},
  pages={7087--7091},
  year={2022},
}

@inproceedings{wang2017tacotron,
  title={{Tacotron: Towards end-to-end speech synthesis}},
  author={Wang, Yuxuan and Skerry-Ryan, RJ and Stanton, Daisy and Wu, Yonghui and Weiss, Ron J and Jaitly, Navdeep and Yang, Zongheng and Xiao, Ying and Chen, Zhifeng and Bengio, Samy and others},
  booktitle={Proc. Interspeech},
  pages={4006--4010},
  year={2017}
}

@inproceedings{loshchilov2017decoupled,
  title={{Decoupled Weight Decay Regularization}},
  author={Ilya Loshchilov and Frank Hutter},
  booktitle={Proc. ICLR},
  year={2017},
}

@inproceedings{smith2019super,
  title={{Super-convergence: Very fast training of neural networks using large learning rates}},
  author={Smith, Leslie N and Topin, Nicholay},
  booktitle={Artificial intelligence and machine learning for multi-domain operations applications},
  volume={11006},
  pages={369--386},
  year={2019},
}

@inproceedings{baba2024t05,
  title={{The T05 system for the VoiceMOS Challenge 2024: Transfer learning from deep image classifier to naturalness MOS prediction of high-quality synthetic speech}},
  author={Baba, Kaito and Nakata, Wataru and Saito, Yuki and Saruwatari, Hiroshi},
  booktitle={Proc. SLT},
  pages={818--824},
  year={2024},
}

@inproceedings{vioni2023investigating,
  title={{Investigating content-aware neural text-to-speech MOS prediction using prosodic and linguistic features}},
  author={Vioni, Alexandra and Maniati, Georgia and Ellinas, Nikolaos and Sung, June Sig and Hwang, Inchul and Chalamandaris, Aimilios and Tsiakoulis, Pirros},
  booktitle={Proc. ICASSP},
  year={2023},
}

@misc{ljspeech17,
  author = {Keith Ito and Linda Johnson},
  title  = {{The LJ Speech Dataset}},
  howpublished = {\url{https://keithito.com/LJ-Speech-Dataset/}},
  year   = {2017}
}

@inproceedings{sculley2010web,
  author    = {David Sculley},
  title     = {{Web-scale k-means clustering}},
  booktitle = {Proc. WWW},
  pages     = {1177--1178},
  year      = {2010},
}

@misc{blizzard2009,
  author       = {Simon King and Vasilis Karaiskos},
  title        = {{The Blizzard Challenge 2009}},
  howpublished = {Blizzard Challenge workshop, University of Edinburgh},
  year         = {2009},
  url          = {https://www.isca-archive.org/blizzard_2009/king09_blizzard.pdf}
}

@article{ba2016layer_corr,
  title   = {{Layer Normalization}},
  author  = {Lei Jimmy Ba and Jamie Ryan Kiros and Geoffrey E. Hinton},
  journal = {CoRR},
  volume  = {abs/1607.06450},
  year    = {2016},
  url     = {https://arxiv.org/abs/1607.06450}
}

@inproceedings{pasad2023comparative,
  title={{Comparative layer-wise analysis of self-supervised speech models}},
  author={Pasad, Ankita and Shi, Bowen and Livescu, Karen},
  booktitle={Proc. ICASSP},
  pages={1--5},
  year={2023},
  organization={IEEE}
}

@incollection{hotelling1992relations,
  title={Relations between two sets of variates},
  author={Hotelling, Harold},
  booktitle={Breakthroughs in statistics: methodology and distribution},
  pages={162--190},
  year={1992},
  publisher={Springer}
}

\end{document}